# High magnetoresistance at room temperature in p-i-n graphene nanoribbons due to band-to-band tunneling effects

Gengchiau Liang[1,a)], S. Bala kumar[1b)], M. B. A. Jalil[1], and S. G. Tan[2]

[1]Department of Electrical and Computer Engineering, National University of Singapore, Singapore 117576, Republic of Singapore
[2]Data Storage Institute, Agency for Science, Technology and Research (A*STAR), DSI Building, 5 Engineering Drive 1, Singapore 117608, Republic of Singapore

A large magnetoresistance effect is obtained at room-temperature by using p-i-n armchair-graphene-nanoribbon (GNR) heterostructures. The key advantage is the virtual elimination of thermal currents due to the presence of band gaps in the contacts. The current at B=0T is greatly decreased while the current at B>0T is relatively large due to the band-to-band tunneling effects, resulting in a high magnetoresistance ratio, even at room-temperature. Moreover, we explore the effects of edge-roughness, length, and width of GNR channels on device performance. An increase in edge-roughness and channel length enhances the magnetoresistance ratio while increased channel width can reduce the operating bias.

Corresponding author:
a) elelg@nus.edu.sg
b) brajahari@gmail.com.

The growth of research activities on graphene-based materials and their applications increases considerably in the last few years due to their unique physical properties and promising device applications. Based on the high carrier mobility[1-3] and recent technological advances which make possible the controlled patterning of graphene nanoribbons (GNR) to induce semiconductor behaviors,[4-6] several types of graphene-based electrical field-effect-transistor (FET) devices[7-11] have been proposed to overcome the physical scaling limitations. In addition, the unique properties of graphene related materials under the magnetic field have also attracted much attention from the scientific and engineering communities. These include the very large magnetoresistance (MR) effects based on the magnetic states of zigzag GNRs with proper spin injections[12-14] and transport gap modulations of armchair GNRs (AGNRs) using the zeroth (n=0) Landau level (LLs) under the magnetic field (B-field).[15-21] In particular, the latter does not require any spin polarized current or precise width of the zigzag GNRs to achieve a high MR, and can be implemented, in principle, in all GNRs. Several theoretical and experimental studies have shown that the very large MR of GNRs can be obtained at low temperature.[15-17] However, at room temperature, MR is significantly suppressed due to the large thermal current at B=0 T.

In this work, we propose a back-gate p-i-n heterostructure AGNR device to prevent the suppression of MR at high temperature. The device consists of a p-type source, an intrinsic channel, and an n-type drain with the Fermi level in the channel aligned to the valence and conduction band-edges of the contacts, as shown in Fig. 1 (a). The potential of the channel is controlled by a back gate, and its profile is kept flat in order to reduce the direct tunneling of electrons. This p-i-n gated structure has been considered as one of the potential candidates for the next generation digital gate[22,23]. Due to the different tunneling paths for the LOW-state and HIGH-state currents, the p-i-n gated structure can considerably reduce the subthreshold slope dominated by the thermionic current, which is limited to 60meV/decade in MOSFETs.

Utilizing the properties of the p-i-n gated structure, as well as the bandgap modulation of AGNRs under the B-field,[17-21] the current at the low conductance state (LOW-state, like OFF-state of a conventional TFET) which is obtained at B=0, is greatly decreased. It is because, unlike the case of AGNR-based devices with metal contacts, the transport of electrons in the p-i-n device is strongly suppressed above (below) the Fermi level in the source (drain) due to the bandgap in the contacts. Hence, only the electrons with energy in between the valence band edge of the source and conduction band edge of the drain are allowed to tunnel directly from source to drain. Since the tunneling width is equal to the length of the channel, as illustrated in Fig. 1(b), a very low conductance is obtained, which we term as the 'LOW-state'. On the other hand, when a magnetic field is applied, the bandgap of the channel AGNR decreases, and hence, its conduction (valence) band overlaps with the valence (conductance) band of the source (drain), c.f., Fig. 1(c). This results in the narrow band-to-band tunneling junction across the source (drain) and channel. Compared to the tunneling width of the LOW-state, the width of the band-to-band tunneling junction is much smaller and hence, the current is increased significantly to yield the HIGH-state.

Furthermore, the AGNR width plays an important role in the bandgap modulation at HIGH-state and the efficiency of thermal current elimination on the LOW-state. The bandgap modulation of the larger width GNR case (i.e. of the channel) is more sensitive to the B-field. Thus, under small B-field, the bandgap ($E_g$) of the channel can be reduced, while that of the source and the drain remains practically unchanged.

Simultaneously, the larger bandgap contacts can suppress the thermal electron more efficiently. Therefore, this heterojunction p-i-n device can be operated under a small B-field applied across the entire device and would enhance MR. On the other hand, due to the tunneling process and narrower contacts, the LOW-state and HIGH-state currents of the p-i-n gated structure are much smaller compared to those of the metal contacts. Hence, the multiple narrow source/drain channels with the same size are used to connect to the channel in order to increase the total currents, as shown in the inset of Fig. 1(b).

To gain fundamental insights into the operation principle, as well as to investigate the performance of these device structure, a full real-space quantum transport simulation based on the non-equilibrium Green's Function (NEGF) approach[24,25] using a simple $\pi$-orbital tight-binding model with edge effect modification[26] is implemented in this work. For the general description of this simulation approach, the readers can refer to Ref. [17]. While the metallic contacts are used in the previous models, here we use semi-infinite AGNR source and drain reservoirs as the contacts. The electronic potential in the channel region is assumed to be a flat well controlled by a back gate voltage, while the depletion width at the contact-channel junctions is estimated using the full-depletion approximation.[27]

Following the operation principle discussed above, we first investigate the current density (J) and the MR effect as a function of the B-field, for the $L_y$=15nm AGNR whose bandgap is 0.07eV with various channel lengths of $L_x$=86, 108 and 130nm under $V_{DS}$=70mV and at room temperature (300 K). The source and drain are implemented by narrower AGNRs with the bandgap of 2.8eV, and 25 strips connecting to the channel part. We assume that there are no interactions between these AGNR source/drain strips. As shown in Fig. 2(a), and (c), it can be found that compared to metal contact with Lx=130nm AGNR as the channel[17], the current J of the p-i-n back-gated AGNR device shows a significant change from B=0 to B=5 T due to the different tunneling processes discussed above. Fig. 2 (e) and (f) further illustrate the tunneling current contributions with the current spectra (J(x,E)) under B=0 and 5 T, respectively. It can be found that at B=0 T, the current is mainly dominated by the direct tunneling from the source to drain through the bandgap of the channel. On the other hand, under B=5T, the bandgap of the channel decreases, resulting in the overlap of band alignment between the source/drain and channel. This overlap allows band-to-band tunneling from the source to the channel and from the channel to the drain. The dramatic decrease in tunneling width results in a large increase in J, thus resulting in a high MR even at room temperature.

Next, we investigate the effect of AGNR channel length on the MR. At LOW-state the tunneling width is equal to the channel length. Therefore, as the channel length increases from $L_x$=86, 108, 130 nm, the tunneling current decreases, resulting in an exponential decreases in the LOW-state current. On the other hand, at HIGH-state, the magnetic field is large enough to reduce the bandgap, allowing band-to-band tunneling between the channel and contacts. Hence, the HIGH-state currents are almost independent of the channel length. As a result, the MR can be enhanced by increasing the channel length.

The MR dependence on the channel width is also explored by varying Ly=15, 9, and 6nm, with 25, 15, 10 strips connecting to the channel, respectively. The intrinsic bandgap of these materials are 0.07eV, 0.17eV, and 0.25eV, and therefore, to align the source and drain properly, the applied $V_{DS}$ is set as 0.07V, 0.17V, and 0.25V. As shown in Fig. 2(b) and (d), we found that as the width is reduced, the J(B=0T) remains the same but J(B=5T) decreases, resulting in a decrease in MR. The former further demonstrates that the LOW-state current is dominated by the direct tunneling current within the energy range between the valence band-edge of the source and the conduction band-edge of the drain. It also indicates that the thermal current, which is the main source of MR degradation observed in AGNR device with metal contacts, has been effectively suppressed by the use of semiconducting contacts and p-i-n tunneling structures. As for the decrease in J(B=5T) as well as MR for the narrower channel widths, this can be attributed to the lower sensitivity of the bandgap of narrower width AGNRs to the applied magnetic field[17]. It also indicates that with a wider channel, the required MR value can be achieved at smaller drain bias and magnetic field.

Lastly, the edge roughness (ER) effects on current and MR are studied using channel AGNR with Lx=130nm and Ly=15nm. The ER is quantified as the percentage of carbon atoms which are dislocated at the edges. For each ER cases, ten different structures are generated by the random removal and addition of edge dimers. As shown in Fig. 3(a), the LOW-state current, J(B=0T), decreases as the ER increases. This is because the transport gaps of the large width AGNRs under B=0T increase with ER[28,29], cf., Fig. 3(c). However, the HIGH-sate current J(B=5T) and transport gaps only varies slightly as the ER increases. The reason can be attributed to the backscattering suppression due to the formation of robust unidirectional edge currents at large B-fields[30]. Thus, we have the interesting and counter-intuitive result that ER effects can actually enhance the MR by as much as two orders of magnitude when the ER is increased from 0% to 25%.

Finally, we would like to note that the purpose of this work is to introduce the operation principle for implementing the unique properties of carbon nano-materials under B-field, and the suppression of thermal currents by using tunneling structures.

Several detailed effects on device performance were not included, such as material selection for heterojunction structures, doping concentrations, ER in the source region and phonon scattering effects. Although the use of nano-width GNRs with large bandgap as the source and drain can suppress thermal currents, the fabrication of these nano-width GNRs presents a practical challenge. We believe that other semiconductor materials can be used to replace the nano-width GNRs in the drain/source contacts, without altering the basic operation principle.[31] However, the effect of band-alignment, band-offset, and the realistic interface at the heterojunctions should be considered carefully in modeling the tunneling mechanism. Furthermore, doping concentrations of the source and the drain can play an important role in controlling the tunnel barrier width and the Fermi level positions. These issues should be studied in the framework of a self-consistent simulation in order to yield details of the device performance optimization. ER in the source-channel region will also lead to changes to the bandgap of the contact, resulting in band alignment between the source and channel,[23] and hence, variation in the operation voltage and magnetic field. Lastly, phonons may play a critical role in increasing the LOW-state currents. The unique properties of graphene result in a suppression of phonon scattering, and thus, ballistic electron transport is assumed in this work. However, it would be very interesting to delve into the physical insight of and the effect of various phonon scatterings on the performance of devices with different heterojunction structures.

In summary, we propose a p-i-n heterostructure AGNR device which exhibits a magnetoresistive (MR) response under an applied magnetic field. The key advantage of the device is the suppression of the LOW-state thermal currents due the bandgap in the source and drain, thus enabling a high MR to be achieved at room temperature. Furthermore, the dependence of the MR on the channel length and width was investigated. We found that, due to the direct source to drain tunneling mechanism, the LOW-state current is reduced exponentially with the increase in channel length increases, thus resulting in a high MR. On the other hand, a wider AGNR channel allows the device to be operated under lower B-field due to the high magnetic field sensitivity of band gap modulation in wide AGNRs. Finally, we found that edge imperfections or roughness can actually enhance the MR ratio, by reducing the LOW-state current while maintaining the HIGH-state current.

*Acknowledgement*—The computations were performed on the cluster of Computational Nanoelectronics and Nano-device Laboratory, National University of Singapore. This work was supported by Agency for Science, Technology and Research, Singapore (A*STAR) under grant number 082-101-0023.

Caption

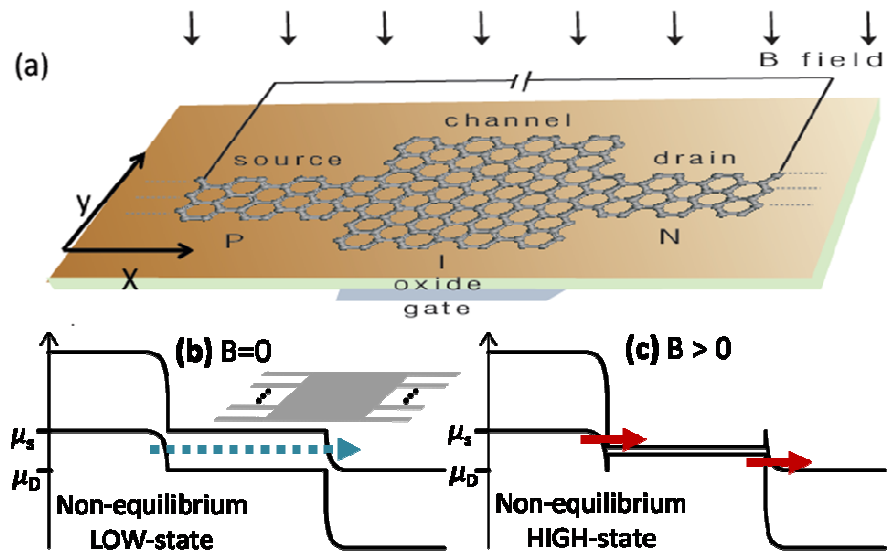

Fig. 1. (color online) (a) A schematic of a back-gate p-i-n heterostructure AGNR device, consisted of the p-type source, intrinsic channel, and n-type drain. The external magnetic field is applied to modulate the bandgap of AGNRs in order to obtain MR. (b) and (c) schematics of band profile under B=0 T and B > 0, identified as LOW-state and HIGH-state, respectively. For the former, the thermal current is eliminate by bandgap of the source and drain and the current is contributed by direct tunneling across the channel, resulting in a very low current. However, for the latter, the bandgap decreases due to the magnetic field, thus creating an overlap between the channel and the source/drain. Therefore, electrons are able to tunnel easily, and contribute a large HIGH-state current, hence resulting in high MR effects of p-i-n AGNR devices even at room temperature. The inset in (b) shows the multiple source-drain channel strips used in the simulation.

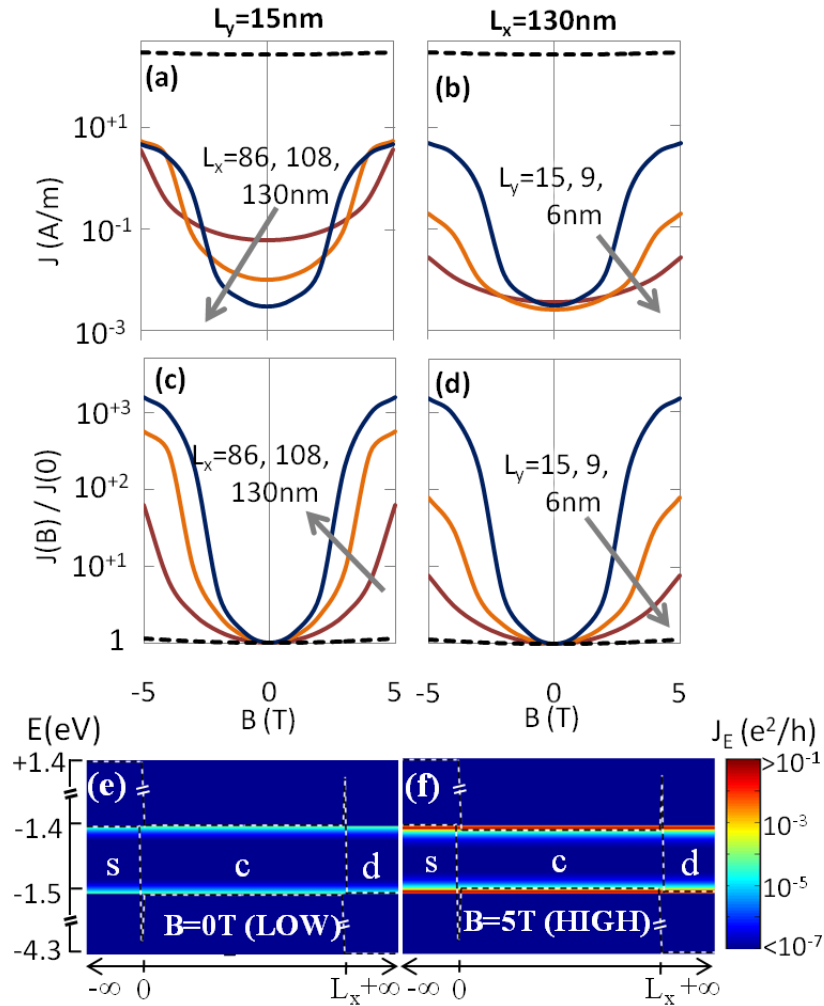

Fig. 2. (color online) Current density, J variation with varying B-field for p-i-n heterostructure AGNR devices with different GNR channels length, $L_x$ when channel width $L_y$=15nm (a), and different GNR width, $L_y$ when channel length, $L_x$=130nm (b). (c) and (d) present the magnetoresistance, J(B)/J(0) variation with varying B-field for different channel structure of (a) and (b), respectively. The dotted lines indicate results of metallic contacts used for source and drain. It can be found that compared to the case of AGNR device with metallic contacts, the current J of p-i-n AGNR devices varies significantly from B=0 to B=5T, resulting in enhancing MR. Current spectra, $J_E(E)$ across the device for (e) LOW-state (B=0T) contributed by the direct tunneling current through the bandgap of the channel, and (f) High-state (B=5T) contributed by band-to-band tunneling current between the source and channel, and between the channel and the drain.

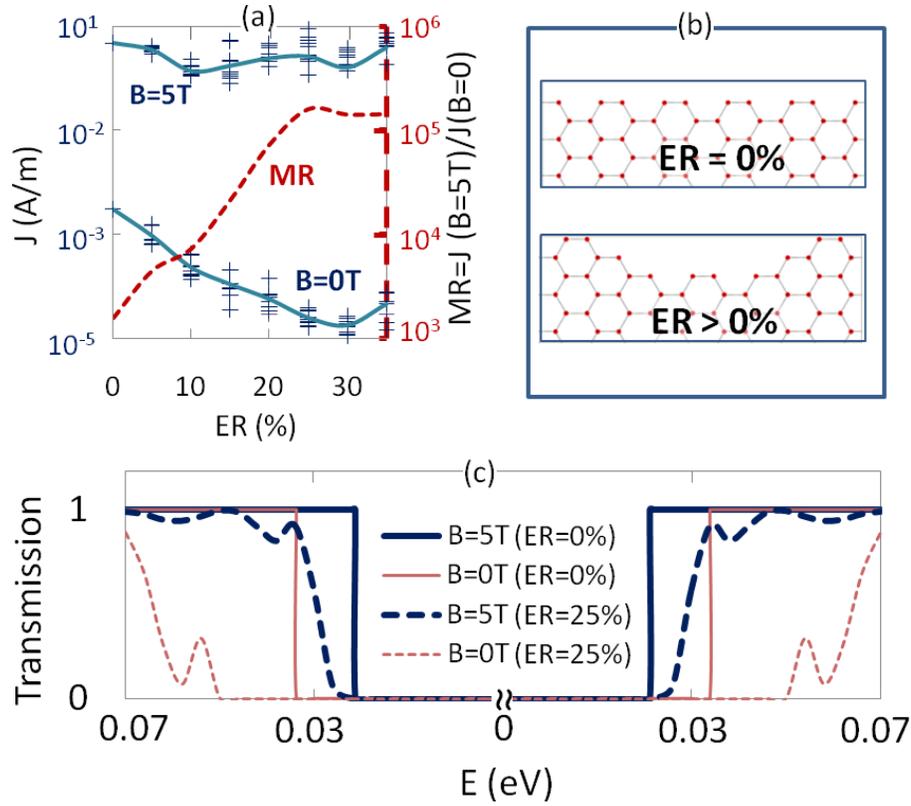

Fig. 3. (color online) (a) Solid curve shows current density, J with increasing channel edge roughness, ER for B=0T and B=5T. For each case, ten different samples were considered. The value for each sample is indicated by '+', and the average value is shown by the solid lines. Dotted curve shows magnetoresistance, J(B=5T)/J(B=0T) with increasing channel edge roughness, ER for B=0T and B=5T. (b) Schematics of ANGR geometries with perfect edge and with roughness along the top edge. (c) The variation of electron transmission at different energy levels, E for B=0T (thick lines) and B=5T (thin lines) when ER=0% (solid lines) and ER=25% (dotted lines). The shift in the curve due to B-field is more significant for the case of ER=25% (dotted lines) compared to the case of ER=0% (solid lines). This ER induced conductance gap is less significant in the presence of a B-field of B=5 T, and thus the bandgap reduction due to B-field is larger for AGNR with ER.